\documentclass[11pt]{article}

\usepackage{epsfig}

\usepackage{amssymb}
\usepackage{graphicx}
\usepackage{color}
\usepackage{subfigure}

\begin{document}

\baselineskip 6mm
\renewcommand{\thefootnote}{\fnsymbol{footnote}}


\def\be{\begin{equation}}\def\ba{\begin{eqnarray}}
\def\ee{\end{equation}}\def\ea{\end{eqnarray}}
\def\no{\nonumber\\}
\def\){\right)}
\def\({\left(}

\def\la{\langle}\def\ra{\rangle}

\begin{titlepage}
\hfill\parbox{5cm} { }

\vspace{25mm}

\begin{center}
{\Large \bf Self-bound dense objects in holographic QCD}
\vskip 1. cm
 {
  Kyung Kiu Kim$^a$\footnote{e-mail : kimkyungkiu@gmail.com },
  Youngman Kim$^{b, c}$\footnote{e-mail : ykim@apctp.org},
  and Yumi Ko$^b$\footnote{e-mail : koyumi@apctp.org} }

\vskip 0.5cm

{\it

$^a\,$ Institute for the Early Universe, Ewha Womans University, Seoul 120-750, Korea \\
$^b\,$Asia Pacific Center
for Theoretical Physics, Pohang, Gyeongbuk 790-784, Korea \\
$^c\,$ Department of Physics, Pohang University
of Science and Technology, Pohang, \\
Gyeongbuk 790-784, Korea \\

}
\end{center}

\thispagestyle{empty}

\vskip2cm


\centerline{\bf ABSTRACT} \vskip 4mm
We study a self-bound dense object in the hard wall model.
We consider a spherically symmetric dense object which is characterized by
its radial density distribution and non-uniform but spherically symmetric chiral condensate.
For this we analytically solve the partial differential equations in the hard wall model  and read off
 the radial coordinate dependence of the density and chiral condensate according to the AdS/CFT  correspondence. We then attempt to describe nucleon density profiles of a few nuclei within our framework
 and observe that the confinement scale changes from a free nucleon to a nucleus.
 We briefly discuss how to include the effect of  higher dimensional operator into our study.
 We finally comment on possible extensions of our work.

\vspace{1cm}

\vspace{2cm}


\end{titlepage}

\renewcommand{\thefootnote}{\arabic{footnote}}
\setcounter{footnote}{0}

\tableofcontents


\section{Introduction}

A self-bound object is a stable bound state in dense matter with strong interactions, {\it i.e.},
not requiring gravity to sustain the bound state unlike a normal neutron star that is gravitationally bound.
An immediate example of such a state is the nucleus.
Nuclei consist of neutrons and protons bound by the nuclear force, or quarks bound by the strong interaction.
The possibility of having strongly bound matter, which would not decay back to ordinary nuclei once formed,
was first discussed by Bodmer~\cite{Bodmer}. A well-known example of this kind is
strange quark matter, in which the energy per
baryon number is less than the mass of the nucleon.
Strange matter has been of great interest ever since the original suggestion
that it could be stable \cite{sMatter} and so  be the ultimate ground state of dense matter.
One of the most important  consequences of the strange matter
hypothesis is the possible existence of strange stars, that is compact stars which are
completely (or almost completely) made of strange quark matter.
 For details on normal neutron stars and
self-bound stars, we refer to \cite{NS_review}, for example.
Another interesting example is the kaon-condensed hypernuclei~\cite{KQball}, which  may be produced in a terrestrial experiment such as
the heavy-ion collision.
The quantities of interest to study the self-bound or gravitationally bound object
 are the structure of the interior and surface of the object, its energy density $\epsilon (\rho)$, and
pressure $P(\rho)$, where $\rho$ denotes the baryon number density.

In this work, we take a first step towards a realistic description of the self-bound object in holographic QCD
based on the AdS/CFT correspondence~\cite{adscft}.
Recent developments of holographic QCD in free space~\cite{free_hQCD} and in dense matter~\cite{dense_hQCD}
are strongly appealing that holographic QCD could be a promising and powerful analytic tool
in understanding non-perturbative nature of quantum chromodynamics (QCD).
It is interesting to notice that holographic neutron stars~\cite{BPV} and gravity dual of giant resonances of heavy nuclei~\cite{KHa}
are recently investigated.
As a first try, we consider an oversimplified self-bound object: a structureless and spherically symmetric object.
The oversimplified object will be characterized by $\rho (r)$ and $\sigma (r)$, where $\rho$ is the baryon number density and
$\sigma$ is quark-antiquark (chiral) condensate.  Here $r$ is for radial coordinate of the spherical object.
For the sake of simplicity, we study the simple object in the hard wall model~\cite{EKSS, PR}.
Since the hard wall model has an infrared (IR) cutoff that is introduced by hand in the extra dimension, it
is naturally plagued by ambiguities in the IR boundary conditions. We will work with several IR boundary conditions  in Section \ref{SBOHW}
and in Appendix \ref{AppA}.
 Then, we introduce a higher dimensional operator~\cite{hD} into the hard wall model to
see the interplay between chiral condensate and the density in Appendix \ref{AppB}.
Though it should remain a long way to go to a claim that we are describing any realistic physical objects, we
comment briefly on the nucleon number density profile in nuclei within our framework. Interestingly, we observe that
 the confinement scale changes going from a free nucleon to a nucleus.
Our hope is that this work would be
a good starting point towards a realistic description of the self-bound or gravitationally bound objects in nature.


\section{Hard wall model with a higher dimensional operator }
 We introduce the hard wall model with $U(N_f)_L \times U(N_f)_R$ gauge symmetry developed in~\cite{EKSS, PR}
and its extended version with  higher dimensional operators~\cite{hD}.
The action of the hard wall model ~\cite{EKSS, PR} is
\begin{eqnarray}
 S_5
 = \int d^4 x \int dz \,\sqrt{g} ~{\rm Tr} \,\left[
-\frac{1}{4g_5^2}\,\(F_L^2+F_R^2\,\)+|DX |^2 +\frac{3}{L^2}\,|X|^2~\right],\label{S5}
\end{eqnarray}
where $D_\mu X = \partial_\mu X - i A_{L\mu}\, X+ i X A_{R\mu}$ and
$A_{L,R}=A^a_{L,R} t^a$ with ${\rm Tr}(t^a t^b)=\frac{1}{2}
\delta^{ab}$. The bulk scalar field is defined by  $X = X_0\,
e^{2i\,\pi^at^a}$, where $X_0 \equiv \langle X \rangle$. The
background metric of the model  is a slice of AdS metric,
\begin{equation}
ds^2 = \frac{1}{z^2}\left( \eta_{\mu\nu}\, dx^{\mu} dx ^{\nu} -dz^2 \right),~~~\epsilon\leq z \leq z_m\, .
\end{equation}
Here $g_5$ is the five dimensional gauge coupling, $g_5^2=12\pi^2/N_c$, and $\epsilon$ ($z_m$) is the UV-cutoff (IR-cutoff).
In~\cite{EKSS, PR} the IR-cutoff $z_m$ is fixed by the $\rho$-meson mass: $1/z_m \simeq 320~{\rm MeV}$.

In the standard AdS/CFT approach, the baryon (quark) number chemical potential is
introduced as a background for the time component of bulk $U(1)$ gauge field $A_t$~\cite{dense_hQCD, DH}.
In the hard wall model~\cite{EKSS, PR}, however, due to specific structures of interaction terms,
the role of the chemical potential is very much limited.
Even though we turn on $A_t$ to introduce the baryon chemical potential, it does not couple to any other bulk fields
in the hard wall model.
A simple way to couple $X_0$ to $A_t$ is to introduce a higher dimensional operator.
As discussed in~\cite{hD}, we can add several higher order terms in the hard wall model.
Among them, there is only one term that could introduce the $U(1)$ chemical potential into the model and that
generates coupled terms involving $A_t$, $X_0$, and other bulk fields. The term is proportional to $ X_0^2F_V^2$, where
$F_V$ is the field strength of the bulk $U(1)$ gauge field.  Here the $U(1)$ gauge group is the vector  subgroup of $U(N_f)_L \times U(N_f)_R$.
Then the modified hard wall model action reads
\ba
S= \int d^5 x \sqrt{g}\, {\rm Tr}\(|DX|^2 + \frac{3}{L^2} |X|^2 - \frac{1}{4 g_5^2} F_V^2 + c_1 |X|^2 F_V^2\). \label{action}
\ea
The equations of motion are
\ba
\nabla^2 X - \frac{3}{L^2} X - c_1 X\,F_V^2 = 0, \label{b1} \no
\nabla_\mu F_V^{\mu\nu} - 4\, g_5^2\,  c_1 \nabla_\mu \(|X|^2 \,F_V^{\mu\nu} \) = 0.\label{b2}
\ea
Since we are going to study a static spherical self-bound object, we take
the metric as
\ba
ds^2 = \frac{L^2}{z^2} \left(\, dt^2 - dr^2 - r^2 \(d\theta^2 + \sin^2 \theta d\phi^2 \,\) - dz^2 \,\right)\, .\label{b3}
\ea
In this background, we solve the equations of motion in Eq.~(\ref{b2}) to calculate the radial coordinate $r$
dependence of $A_t$ and $X_0$.
{}From $A_t$ we can read off $\rho(r)$, baryon number density distribution.
We can also study the $r$ dependence of the chiral condensate that is basically the boundary value of $X_0$.
Expected behavior is that at the center of the object, $r=0$, the baryon number density will be large,
while the chiral condensate will be very small.
At the surface of the object,  the baryon number density will be (much) smaller than the central value,
and the chiral condensate will be larger than the center value.

\section{Self-bound objects in the hard wall model \label{SBOHW}}
We first consider the case with $c_1$=0.
In this case the baryon number density and chiral condensate are quite independent from each other
 since $X_0$ and $A_t$ do not couple to each other, which is not the case
in any realistic dense matter; the chiral condensate decreases with increasing baryon number density.
However, we can realize this situation in this hard wall model by imposing suitable boundary conditions
at the center and on the surface of our spherical object. This is because the chiral condensate and baryon number density
are nothing but integration constants of equations of motion for $X_0$ and $A_t$.
Note that in the hard wall model~\cite{EKSS, PR} nonzero value of the chiral condensate was from the boundary
condition not from a scalar potential that supports nonzero vacuum expectation value of $X$.
This feature of the hard wall model was improved in \cite{PR2}, where a scalar potential $V(X)$ is introduced to obtain a nonzero value of chiral condensate by minimizing an action.
Since we are interested in the baryon number density and chiral condensate, we turn on $A_t$ and consider the vacuum expectation
value of $X$.
The equations of motion are then
\ba
\partial_z^2 A_t- \frac{\partial_z A_t}{z} + \partial_r^2 A_t + \frac{2 \partial_r A_t}{r} = 0 \, ,\nonumber\\
\partial_z^2 v - \frac{3 \partial_z\,v}{z} + \frac{3 v}{z^2}+\partial_r^2 v +\frac{2 \partial_r v}{r} = 0\, ,\label{b4}
\ea
where $X_0 (r,z)= {v(r,z)}/{2}$. Noticing that the equations are separable, we
decompose the bulk fields as
\be
A_t ( r, z) = A(r) B(z), \quad v(r, z) = C(r) D(z)\, . \label{b5}
\ee
Then we obtain four decoupled second order differential equations that can be solved analytically.
We first consider the gauge field.
The equation for $A(r)$ is given by
\ba
{\ddot A}(r) + \frac{2}{r} {\dot A}(r) - l^2 A(r)=0\, , \label{b9}
\ea
where we use `` $\cdot$ '' for a derivative with respect to $r$. Then the solutions read
\ba
&&A(r) = a_0 - \frac{a_1}{r}\,,  \quad l=0,\nonumber\\
&&A(r) = a_3 \frac{e^{-l r}}{r} + a_4\, \frac{e^{lr}}{ r}\,, \quad l= {\rm const.}\label{b10}
\ea
For $B(z)$ we have
\ba
B''(z) - \frac{1}{z} B'(z) + l^2 B(z) = 0\, .\label{b6}
\ea
Here  `` $'$ ''  denotes a derivative with respect to $z$ and
\ba
&&B(z) = b_0+ \frac{b_1}{2} z^2 ,  \quad l=0\,,\nonumber\\
&&B(z) = b_3\, z J_1 (lz) + b_4\, z Y_1 (l z ), \quad l= {\rm const.} \label{b7}
\ea
Collecting the solutions, we write the general solution
\ba
A_t ( r, z) =  \left( a_0 - \frac{a_1}{r}\right)\left( b_0 + \frac{b_1}{2} z^2 \right) +\sum_l  z J_1 (lz)\( \alpha_l \frac{\cosh (lr)}{r} + \beta_l \frac{\sinh(lr)}{r} \)\, .\label{b11}
\ea
The equations of motion for the scalar field and corresponding solutions are
\ba
&&{\ddot C}(r) + \frac{2}{r} {\dot C}(r)  - k^2 C(r) = 0,\no \label{b14}
&&C(r) = c_0 - \frac{c_1}{r}, \quad k=0 \,,\nonumber\\
&&C(r) = c_3 \frac{e^{-kr}}{r} + c_4\frac{e^{k r}}{r }, \quad k={\rm const.} \, ,\label{b15}
\ea
and
\ba
&& D(z)'' - \frac{3}{z} D(z)' + \left( \frac{3}{z^2} + k^2 \right) D(z) = 0 \, ,\no \label{b12}
&&D(z) = d_0 z + d_1 z^3, \quad k=0\,, \nonumber\\
&&C(z) = d_3 z^2 J_1 (kz) + d_4 z^2 Y_1 (kz), \quad k={\rm const.} \, \label{b13}
\ea
The general solution, which is regular at $r=0$, takes a form
\ba
v(r, z) =d_0 z+  d_1 z^3 + \sum_k \delta_k \left(\frac{\sinh (kr)}{r} \right) z^2 J_1 (kz)\, . \label{b16}
\ea
Expanding the solution near the boundary $z=0$, we confirm that the solution follows what it should be from the AdS/CFT correspondence as
\ba
v(r,z) \sim d_0 z+ \left(d_1 + \sum_k \delta_k \frac{k \sinh(kr)}{r}\right) z^3  + {\cal O}(z^5) \, .\label{b17}
\ea
Since $v$ or $\la X\ra$ is dual to a quark bi-linear operator of the boundary gauge theory, $d_0$ corresponds to the quark mass
up to a constant. For simplicity, we work in the chiral limit, where the quark mass is zero.
Now we are about to impose boundary conditions to fix integration constants above.
A well-known plague in the hard wall model is an ambiguity in the IR boundary at $z=z_m$.
At this moment we do not know how to avoid this problem, and so we make some choices of the IR
boundary condition below and in Appendix \ref{AppA}.

\subsection{ $A_t (r, z_m) = 0$ \label{BCA1}}
\begin{figure}[t]
\center
\includegraphics[width=8cm]{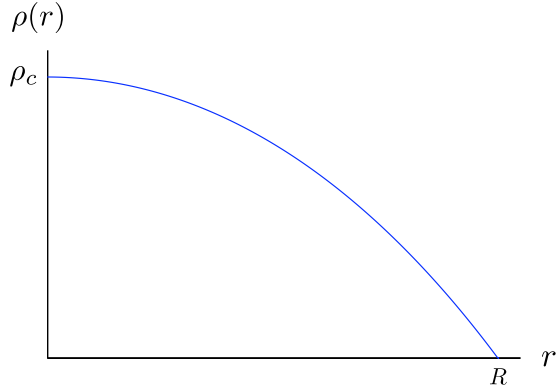}
\caption{Density distribution given in Eq.~(\ref{c11}) }\label{bb}
\end{figure}

The general solutions in the chiral limit are
\ba
A_t ( r, z) \!\!\!&=&\!\!\! \left( a_0 - \frac{a_1}{r}\right) \left( b_0 + \frac{b_1}{2} z^2 \right) +\sum_l  z J_1 (lz) \left(\alpha_l \frac{e^{-l r}}{r} + \beta_l\, \frac{e^{lr}}{ r} \right) \label{c1}\, ,\\
v(r, z) \!\!\!&=&\!\!\! c_0 z^3 +\sum_k \delta_k \left(\frac{\sinh (kr)}{r} \right) z^2 J_1 (kz)\,. \label{c2}
\ea
To ensure the regularity of $A_t$ at $r=0$, we impose  $a_1 =0$ and $\alpha_l = -\beta_l$. Then the solution becomes
\ba
A_t ( r, z) \!\!\!&=&\!\!\! \left( \mu + \frac{\tilde b_1}{2} z^2 \right) +\sum_l  z J_1 (lz) \left(\beta_l\, \frac{\sinh(lr)}{ r} \right) \, ,\label{c3}
\ea
where $\mu=a_0b_0$, which is identified with the chemical potential of the U(1) charge, and $\tilde b_1=a_0 b_1$.
If we choose an IR boundary condition as $A_t(r,z_{m}) = 0$ and require that two terms on the right hand side of Eq.~(\ref{c3})
vanish separately, we obtain
\ba
A_t ( r, z) \!\!\!&=&\!\!\! \left( \mu - \frac{\mu}{z_m^2} z^2 \right) +\sum_n z J_1 \left(\frac{w_n}{z_{m}}z \right) \left(\beta_n\, \frac{\sinh(\frac{w_n}{z_m}r)}{ r} \right), \label{c4}
\ea
where $w_n$ is n-th zero of $J_1(l z)$. Now we can read off the baryon number density from this solution.
At the boundary $z=0$, $A_t$ should be
\ba
A_t ( r, z) \sim \mu + \rho(r) \,z^2 + \cdots\, , \label{c5}
\ea
where
\ba
\rho(r) = -\frac{\mu}{z_m^2}+ \sum_n \frac{w_n \beta_n}{2 z_m\,r}\sinh\left(\frac{w_n}{z_m}r\right) \label{c6}
\ea
Now we impose two boundary conditions at the center $r=0$ and at the surface $r=R$ such that $\rho(0) = \rho_c$ and $\rho(R) = \rho_R$. Then we have
\ba
\rho_c  \!\!\!&=&\!\!\! -\frac{\mu}{z_m^2}+ \sum_n  \frac{\beta_n}{2}\, \left(\frac{w_n}{z_m}\right)^2,\label{c7} \\
\rho_R \!\!\!&=&\!\!\! -\frac{\mu}{z_m^2}+ \sum_n   \left(\frac{w_n}{2z_m} \right) \left(\beta_n\, \frac{\sinh(\frac{w_n}{z_m}R)}{ R} \right).\label{c8}
\ea
Here $\rho_c$, $\rho_R$ and $R$ are external parameters whose values are supposed to vary from objects to objects.
For instance, suppose we are interested in the nucleon density as a function of $r$, a radius of a nucleus which contains $A$ nucleons distributed
 inside a sphere with a radius $R$, then we can choose
 \ba
 \rho_c\sim 0.17~ {\rm fm}^{-3},~ ~ \rho_R\sim 0 ~~{\rm with} ~~ R\sim 1.2\, A^{1/3}\, {\rm fm} \,.
 \ea
If we consider the lowest mode only\footnote{ We do not consider higher modes here since they are expected to have higher energy.}, $n=1$ from (\ref{c7}) and (\ref{c8}), we obtain
\ba
\mu \!\!\!&=&\!\!\!  - \rho_R z_m^2 +  \frac{z_m\,w_1\,\beta_1}{2 R} \sinh \left(\frac{w_1 R}{z_m}\right), \label{c9}\\
\beta_1 \!\!\!&=&\!\!\! 2 \(\rho_c - \rho_R\)\left( \,\frac{w_1^2}{z_m^2}  - \frac{w_1}{z_m R} \sinh \left(\frac{w_1}{z_m} R\right)\right)^{\!\!-1}. \label{c10}
\ea
Note that the values of $\mu$ and $\beta_1$ are determined by $w_1$, $\rho_c$, $R$, $\rho_R$, and $z_m$. As a result, $\mu$ is a $r$-independent
constant and $\rho(r)$ is given by
\ba
\rho(r) = -\frac{\mu}{z_m^2}+\frac{w_1 \beta_1}{2 z_m\,r}\sinh\left(\frac{w_1}{z_m}r\right) \label{c11}.
\ea
In Fig.~\ref{bb}, we plot the density as a function of $r$,where we take $\rho_R=0$.

\subsection{ $v'(r, z_m )=0$}
\begin{figure}[ht]
\center
\includegraphics[width=8cm]{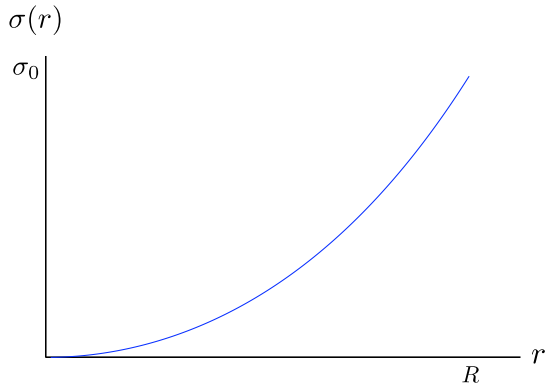}
\caption{ Chiral condensate as a function of $r$ }\label{f2}
\end{figure}

Similarly we obtain
\ba
v(r, z) =\sum_n \delta_n \left(\frac{\sinh (\frac{x_n}{z_m} r)}{r} \right) z^2 J_1 \(\frac{x_n}{z_m} z\) \, ,\label{c14}
\ea
and
\ba
\sigma(r) = \sum_n \frac{x_n \delta_n}{2 z_m r} \,\sinh\(\frac{x_n}{z_m}\, r\) \, .\label{c16}
\ea
Now we impose the following boundary conditions: $\sigma(r) = 0$ at $r=0$ and $\sigma(r) = \sigma_0$ at $r=R$, where $\sigma_0$ is
the value of the chiral condensate in free space. The assumption for this choice is that at the center of a spherical object the baryon
number density is high enough to induce chiral symmetry restoration and at the surface the density is almost zero.
Then we have
\ba
0\!\!\!&=&\!\!\!\sum_n \frac{1}{2} \frac{x_n^2}{z_m^2} \delta_n \label{c17}\, ,\\
\sigma_0 \!\!\!&=&\!\!\! \frac{x_n \delta_n}{2 z_m R} \sinh\(\frac{x_n}{z_m} R\)\label{c18} \, .
\ea
As a simplest nontrivial case we consider the lowest two modes in the Bessel function,
\ba
x_1^2 \,\delta_1 = - x_2^2\, \delta_2, \quad  \delta_2 = - \frac{x_1^2}{x_2^2} \,\delta_1, \label{c19}
\ea
and end up with
\ba
v(r,z) = \delta_1 \(\frac{\sinh\(\frac{x_1}{z_m}\,r\)}{r}\)  z^2 J_1 \(\frac{x_1}{z_m} z\) - \frac{x_1^2}{x_2^2}\( \frac{\sinh\(\frac{x_2}{z_m}\,r\)}{r} \) z^2 J_1 \(\frac{x_2}{z_m} z\)\label{c20}.
\ea
As we expand above equation near $z=0$,
\ba
\sigma(r) = \frac{\delta_1 x_1}{2\, z_m}\( \frac{\sinh\(\frac{x_1}{z_m}r\)}{r} \)  -\frac{\delta_1 x_1^2}{2\, x_2 z_m}\( \frac{\sinh\(\frac{x_2}{z_m}r\)}{r} \),  \label{c21}
\ea
where $\delta_1$ is determined from Eq.~(\ref{c18}) as follows:
\ba
\delta_1 = 2 \sigma_0 \left[ \frac{x_1}{z_m}\( \frac{\sinh\(\frac{x_1}{z_m}R\)}{R} \)  -\frac{x_1^2}{x_2 z_m}\( \frac{\sinh\(\frac{x_2}{z_m}R\)}{R} \) \right]^{-1}  .\label{c23}
\ea
Now we determined every parameters and $\sigma(r)$ in (\ref{c21}) is shown in Fig.~\ref{f2}.

\subsection{Density distribution in nuclei }
\begin{figure}[t]
\center
\includegraphics[width=9cm]{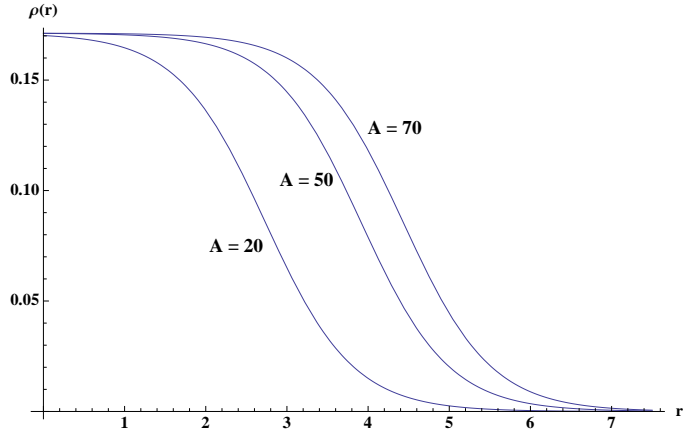}
\caption{Nucleon density as a function of the distance to the center of the nucleus}\label{f3}
\end{figure}

Nucleon number density $\rho(r)$ in a nucleus can be compiled from
the charge distribution $\rho_{ch}(r)$ in nuclei which is measured by elastic electron scattering experiments.
With the assumption that neutron and proton distributions in nuclei are the same,
we can obtain the number density profile of a nucleus having $A$ nucleons and $Z$ protons as
\ba
\rho(r)=\frac{A}{Z}\rho_{ch}\, ,
\ea
where $\rho_{ch}$ is the charge distribution of a nucleus.
A very schematic view of the nucleon density profile of several nuclei are shown in Fig.~\ref{f3}. To draw this we use the following formula which is valid for $A>20$,
\ba
\rho(r)=\frac{\rho_0}{1+e^{(r-R_{1/2})/a}}\, ,\label{rhoexp}
\ea
where $\rho_0$ is the normal nuclear density, $a=0.54~{\rm fm}$, and $R_{1/2}=R_s -0.89 A^{-1/3}$ fm with $R_s=1.128 A^{1/3}$ fm~\cite{NPnut}.
Here the radius of a nucleus is given by $R\simeq r_0 A^{1/3}$, where $r_0\sim 1.2$ fm. This relation is easy to understand. Suppose we have A nucleons of radius
$r$ inside a nucleus of radius $R$ with assumption that both of them are
hard spheres, then we can write
\ba
A\simeq \frac{\frac{4}{3}\pi R^3}{\frac{4}{3}\pi r^3} \, .
\ea
{}From this one can have $R\simeq r_0A^{1/3}$, where $r_0$ is a constant.
If we assume that nucleons are uniformly distributed inside a nucleus, the
uniform number density proportional to
 $ A/R^3$ is independent of the nucleon number A.

One remarkable point in the figure is that the plateau regime, in which the number density is about the normal nuclear matter density,
becomes wider with increasing $A$.

\begin{figure}[ht]
\center
\includegraphics[width=9cm]{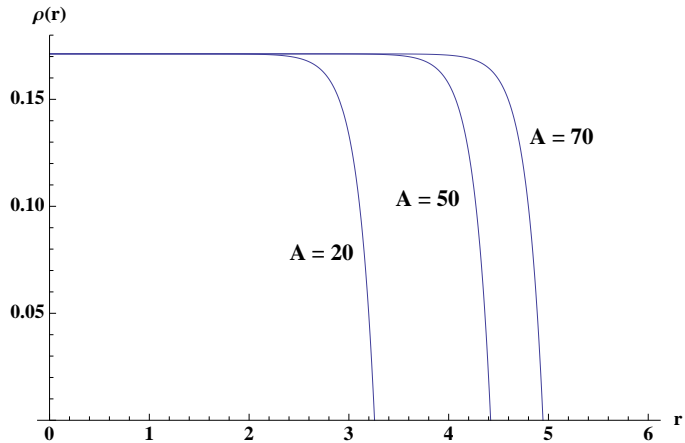}
\caption{Nucleon density distribution in nuclei obtained from holographic QCD with  $z_m$ fixed in the hard wall model, where $1/z_m\sim 320$ MeV.}\label{freezm}
\end{figure}

\begin{figure}[ht]
\center
\includegraphics[width=9cm]{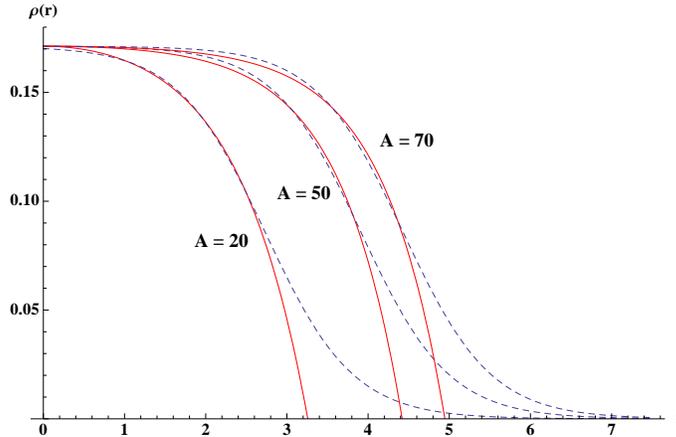}
\caption{Nucleon density as a function of the distance to the center of the nucleus obtained from holographic QCD:  $1/z_m\sim 72.8$ MeV for $A= 20$, and $1/z_m\sim 79.0$ MeV for $A= 50$, and $1/z_m\sim 78.5$ MeV for $A= 70$.}\label{f4}
\end{figure}

To show some applicability of our study to a physical system, we explore the nucleon number density distribution in
nuclei using the result in section \ref{BCA1}.
Since our approach itself is simple and  we only consider strong interactions, we are not able to
explain fine structure of the distribution, but only a global feature of it.
In our approach we have no parameter corresponding to the nucleon number A, which may be done with Chern-Simons term.
To incorporate the A-dependence, we use the relation $R\simeq r_0 A^{1/3}$, and so
the size of a nucleus is an input in our study.
We first show our result with $z_m$ fixed in the hard wall model, $1/z_m\sim 320$ MeV, in Fig.~\ref{freezm}, which looks quite different from
Fig.~\ref{f3}.
We then plot a few density profiles with different IR cutoffs in Fig.~\ref{f4}.
Here we use a different value of $z_m$ from the hard wall model
to simulate the fact that the plateau in the density profile expands with $A$  in Fig.~\ref{f4}; $1/z_m\sim 72.8$ MeV for $A= 20$, and $1/z_m\sim 79.0$ MeV for $A= 50$, and $1/z_m\sim 78.5$ MeV for $A= 70$.
 In Table 1, we list several values of $A$ with corresponding $1/z_m$. Here $z_m$ is fixed such that for each $A$, our result reproduces the plateau regime of
the nucleon number density profile shown in Fig.~\ref{f3}. To have a criterion for the closeness of our results to Fig.~\ref{f3}, we consider the root mean square (rms)
$\sqrt{\frac{1}{r_0} \int_0^{r_0} |\rho_1(r)-\rho_2(r)|^2dr}$, where $\rho_1(r)$ is the profile in Eq.~(\ref{rhoexp}) and $\rho_2(r)$ is our result.
 Here $r_0$ is chosen as the largest value of $r$ satisfying $\rho_1(r)=\rho_2(r)$. For each $A$ we evaluate the rms as a function of $z_m$ to fix the value of $z_m$
that renders the lowest value of the rms.
We note here that typical errors involved in holographic QCD studies are $(10-30)\%$, and so we should not
seriously take the $z_m$-dependence of $A$ in the table also in Fig.~\ref{f4};
at best, we may say that the value of $z_m$ for a nucleus can be different that for a free nucleon.
As one can see from Fig.~\ref{f4}, our results with  $z_m$ different from the hard wall model is much closer to  Fig.~\ref{f3} when we focus on the plateau.
{}If we take our result seriously we can say that varying the nucleon number $A$
is related to a change in our IR cutoff $z_m$.
Note that $1/z_m$ in the hard wall model was introduced to ensure the confinement,  and so it is intimately related to
a confining scale.  Therefore,  we arrive at a qualitative conclusion that the
confinement scale\footnote{Here we assume that the confinement scale of mesons and baryons are the same.}, $\sim 1/z_m$ in our approach, changes from a nucleon to a nucleus.
Though it may not be easy to confirm if our finding is true through any experiments, it is interesting to note the
claim made to explain the European Muon Collaboration (EMC) effect; the quark confinement scale changes in nuclei compared to that of free nucleon
~\cite{EMC}.

\begin{center}
\noindent Table 1 : The nucleon number $A$ and the IR cutoff $z_m$.
\vskip 0.3cm
\begin{tabular}{|c||c|c|c|c|c|c|}
\hline
 $A$ & $1/z_m$\\
\hline
20 & 72.8 MeV \\
\hline
30 & 77.5 MeV\\
\hline
50 & 79.0 MeV\\
\hline
70 & 78.5 MeV\\
\hline
100 & 77.0 MeV\\
\hline
\end{tabular}  \\
\vspace{0.3cm}
\end{center}

\section{Discussion}
We studied a self-bound dense object in the hard wall model.
For simplicity, we considered a spherically symmetric dense object which is characterized by
its radial density distribution and non-uniform chiral condensate.
We analytically solve the partial differential equations of motion derived in the hard wall model and read off
 the radial coordinate dependence of the density and chiral condensate using the AdS/CFT dictionary.
 Since we have an ambiguity in the IR boundary condition, we tried several different boundary conditions.
Though our model study might be still far from any realistic physical systems,
 we tried to understand nucleon density profile of a nucleus within our framework.
 Interestingly, we observed that
 the confinement scale, $1/z_m$ in our model, might change from a free nucleon $\sim 300$ MeV to a nucleus $\sim 100$ MeV.
We incorporate the higher dimensional operator into our analysis in Appendix \ref{AppB}.

Now we discuss how come the self-bound object forms in the hard wall model.
Generally speaking, the binding is due to the strong interaction of gluons or gauge fields of a boundary gauge theory
which is encoded in the dual gravity background, the AdS background in our case.
In our analysis, we did not analyze the stability of the bound object.
Therefore it is still not clear how we can have such an object in the hard wall model, even though
we believe that the binding would be supported by the strong interactions or dual gravity background.
In fact in our study the binding is realized by the boundary conditions in the AdS background.
To visualize this, we take the chiral condensate in the hard wall model as an example.
As pointed out in Ref.~\cite{PR2}, no scalar potential $V(X_0)$ to support the chiral condensate exists
in the hard wall model \cite{EKSS, PR}.
Then, the nonzero chiral condensate was achieved by assigning nonzero value to
an integration constant which corresponds chiral condensate according to the AdS/CFT dictionary.
The value of the chiral condensate is determined by observables such as the $a_1$ mass or pion decay constant.
In our case we also make use of the fact that the chemical potential, number density and chiral condensate
are related to integration constants to assign them suitable values.

Certainly our present work is leaving many things behind.   We didn't properly address the issue of stability of our solutions or our bound objects.
Therefore there exists a possibility to have stable solutions other than the ones obtained in our study.
Analysis on the stability of the bound object in a well-defined model should follow.
The ambiguity in the IR boundary condition needs to be removed or to be well under control to obtain any conclusive results.
These two issues are relegated to our future study in a D3/D7 model~\cite{K3T}.
One immediate generalization of our work might be to study nuclear collective excitation modes.
Possible connection between our bound objects with chiral nuclei in chiral liquids \cite{Lynn} might be interesting to dig into.
One may also consider Q-stars~\cite{Qstar} or neutron stars; in the case of gravitationally bound neutron star one has to introduce the boundary gravity.

\section*{Acknowledgements}
YKim thanks Koji Hashimoto for helpful discussion. This work was supported in part by WCU Grant No. R32-2008-000-101300 (K. Kim).
YKim acknowledges the Max Planck Society (MPG),
the Korea Ministry of Education, Science, Technology (MEST), Gyeongsangbuk-Do and Pohang City for the support
of the Independent Junior Research Group at the Asia Pacific Center for
Theoretical Physics (APCTP).

\appendix
\section{More on IR boundary conditions \label{AppA}}
In this appendix, we present some more choices of the IR boundary conditions.

\subsection{ $A_t'( r, z_m) = 0$}
\ba
A_t'(r,z_m) = b_1 z_m + \sum_l  \beta_l \(\frac{\sinh(lr)}{r}\) \(z J_1 (l z_m)\)'. \label{c24}
\ea
In order to satisfy a boundary condition $A_t'(r, z_m)=0$ we set  $b_1 = 0$,  and  $ \(z J_1 (y_n)\)' = 0$ where $y_n = l z_m$. Then we have
\ba
A_t (r,t)=\mu+\sum_n \beta_n \(\frac{\sinh\(\frac{y_n}{z_m}r\)}{r}\) z J_1 \(\frac{y_n}{z_m}z\) .\label{c25}
\ea
Expanding this equation near $z\rightarrow0$,
\ba
A_t (r,z) = \mu + \frac{1}{2} \sum_n \frac{y_n}{z_m} \beta_n  \(\frac{\sinh\(\frac{y_n}{z_m}r\)}{r}\)  z^2 +\cdots, \label{c26}
\ea
the density can be read off as
\ba
\rho(r) = \frac{1}{2} \sum_n \frac{y_n}{z_m} \beta_n  \(\frac{\sinh\(\frac{y_n}{z_m}r\)}{r}\)  .\label{c27}
\ea
We impose the boundary conditions at $r=0,~R$ as $\rho(0) = \rho_c$ and $\rho(R) = \rho_R$
\ba
\rho_c \!\!\!&=&\!\!\!  \frac{1}{2} \sum_n \frac{y_n^2}{z_m^2} \beta_n, \label{c28} \\
\rho_R \!\!\!&=&\!\!\!  \frac{1}{2} \sum_n \frac{y_n}{z_m} \beta_n  \(\frac{\sinh\(\frac{y_n}{z_m}R\)}{R}\). \label{c29}
\ea
If we consider only the lowest mode for $n$ from Eq.~(\ref{c28}) we have
\ba
\beta_1 = 2 \rho_c \frac{z_m^2}{y_1^2}, \label{c30}
\ea
and $\rho_R$ is determined as
\ba
\rho_R \!\!\!&=&\!\!\!\frac{ \rho_c \,z_m}{y_1 R}\sinh\(\frac{y_1}{z_m}R\) \label{c31}\, .
\ea

\subsection{ $v(r, z_m )=0$}
Now let us consider the scalar solution.
At $z=z_m$, we have
\ba
v(r, z_m) =c_0 z_m^3 +\sum_k \delta_k \left(\frac{\sinh (kr)}{r} \right) z_m^2 J_1 (kz_m) \, .\label{c32}
\ea
Again we make a simple choice, $c_0 = 0$ and $J_1 (h_n) =0$ where $h_n = k z_m$,
 to meet the IR boundary condition $v(r, z_m)=0$
\ba
v(r, z) =\sum_n \delta_n \left(\frac{\sinh (\frac{h_n}{z_m} r)}{r} \right) z^2 J_1 \( \frac{h_n}{z_m}z\) .\label{c33}
\ea
Near the boundary $z\rightarrow0$, we can read off the chiral condensate as a function of $r$,
\ba
\sigma(r) = \frac{1}{2}\sum_n \delta_n \left(\frac{\sinh (\frac{h_n}{z_m} r)}{r} \right) \frac{h_n}{z_m} \label{c34}\, .
\ea

\section{Analysis with higher dimensional operator \label{AppB}}

In this Appendix, we take the hard wall model with a higher dimensional operator
and present basic equations that could be numerically solved.
We first turn off the $r$-dependence to discuss the uniform distribution.
Then we consider a non-uniform case.
Here we will not attempt to solve equations of motion numerically, partly because
due to the ambiguity in the IR boundary condition the numerical solution may not mean much.
In addition, it is expected that the effect of the higher dimensional operator will not change the result obtained in Sec. \ref{SBOHW}
significantly.

\subsection{Homogeneous distribution}
The equations of motion can be written by
\begin{eqnarray}
&&v'' + {\ddot v}- \frac{3}{z}v' + \frac{2}{r} {\dot v}+ \( \frac{3}{z^2} - \frac{2 \,c_1\,z^2 }{L^2} \(\,A_t'^2 + {\dot A_t}^2\)\) v=0 \label{d5}\,, \\&&
A_t'' + {\ddot A_t} - \( \frac{1}{z} + \frac{2 \,c_1\, g_5^2\,v\,v'}{1 - c_1 g_5^2 v^2}\) A_t' + \( \frac{2}{r} - \frac{2\, c_1\,g_5^2\, v\, {\dot v} }{1-c_1\, g_5^2 v^2}\) {\dot A_t} = 0 \label{d6}~,
\end{eqnarray}
where ``$ ~\dot{} ~$" and ``$ ~{}' ~$" denote derivatives with respect to $r$ and $z$. Since these are not separable, it is not easy to solve them analytically.  Here we consider only z dependent solution. In this case, the equations of motion become simpler as follows:
\ba
&&A_t''(z) - \( \frac{1}{z}+\frac{2\, c_1\, g_5^2 \, v(z)\, v'(z)}{1- c_1\, g_5^2\,  v(z)^2}\)\,A_t'(z) = 0 \,,\label{d1}\\&&
v''(z) - \frac{3}{z} v'(z) + \(\frac{3}{z^2} - \frac{2 \,c_1\,z^2\, A_t'(z)^2}{L^2}\) = 0\label{d2}~.
\ea
From the above, one can easily see that the equation for the gauge field is integrable. Thus we can write down  the gauge field with $v(z)$ as
\ba
A_t (z) = \mu - \rho \int_0^z \frac{ x}{1 - c_1\,g_5^2\, v(x)^2}\, dx ~~,\label{d3}
\ea
where the chemical potential $\mu$ is given by our IR boundary condition, then the chemical potential is
\begin{eqnarray}
\mu=\rho \int_0^{z_m} \frac{ x}{1 - c_1\,g_5^2\, v(x)^2}\, dx~.
\end{eqnarray}
Putting it into Eq.~(\ref{d2}), we get
\ba
v''(z) - \frac{3}{z} v'(z) + \(\frac{3}{z^2} - \frac{2 \,c_1\, \rho \,z^4}{L^2(1-c_1\,g_5^2 \,v(z)^2)^2}\) v(z) = 0\label{d4}~.
\ea
This is the only equation that we have to solve in this case.

\subsection{Perturbation with higher dimensional operator}
Now we assume that the correction from the higher dimensional operators is very small since $c_1$ is small~\cite{hD}.
Therefore we can take $c_1$ as an expansion parameter in perturbation. The equations of motion
   Eq.~($\ref{b2}$) can be rewritten more explicitly as
\begin{eqnarray}
&&\(\partial_i^2 + z^3 \,\partial_z  \frac{1}{z^3} + \frac{3}{z^2} \) \,v = \frac{c_1}{4 z^2}\, v \,\( F_{\mu\nu}\)^2 ,\\&&\(\partial_i^2 + z \partial_z \frac{1}{z} \partial_z \) A_t = - \frac{c_1\, g_5^2\,}{2 z^4} \nabla_\mu \(v^2 F^{\mu t}\,\)~,
\end{eqnarray}
where we have used a metric $ds^2 = \frac{dt^2 -\delta_{ij} dx^i dx^j  -dz^2}{z^2}$ for convenience. Expanding $v$ and $A_t$  as $v \sim v^{(0)} + c_1 v^{(1)} + {\cal O}(c_1^2)$ and $A_t \sim  A^{(0)}_t  + c_1 A^{(1)}_t + {\cal O}(c_1^2)\,$, the corrections can be expressed
in terms of  Green's functions as
\begin{eqnarray}
&&v^{(1)}(z,\vec x\,) = \int dz' d\vec x\,' G^v (z,\vec x\, ; z' \vec x\,'\,)\,\frac{1}{4 z^2}\, v^{(0)} ( \,F^{(0)}_{\mu\nu}\,)^2\, (z',\vec x\,')\,,\label{correction}\\\nonumber&&
A_t^{(1)}(z,\vec x\,) =-\int dz' d\vec x\,' G^A (z,\vec x \,; z' \vec x\,'\,) \,\frac{ g_5^2}{2 z^4} \nabla_\mu ((v^{(0)})^2\, F^{(0)\mu t}\,)(z',\vec x\,'\,)~.
\end{eqnarray}
Thus we have to find the Green's functions in order to obtain  the corrections. It can be done easily by considering the boundary conditions carefully. Assuming a Dirichlet boundary condition for $A_t$ and Neumann boundary condition for $v$ at IR cutoff, we obtain the Green's functions as follows:
\begin{eqnarray}
G^v (z,\vec x\,; z',\vec x\,'\,) \!\!\!&=&\!\!\! -\sum_{n=1}^\infty \frac{z^2}{4\pi N^v_n z'} \,J_1\(\frac{x_n}{z_m} z\)J_1\(\frac{x_n}{z_m} z'\) \frac{\exp \(- \frac{x_n}{z_m} |\vec x - \vec x'|\)}{ |\vec x - \vec x'|}\,,\\G^A (z,\vec x\,; z',\vec x\,'\,) \!\!\!&=&\!\!\! -\sum_{n=1}^\infty \frac{z}{4\pi N^A_n } \,J_1\(\frac{w_n}{z_m} z\)J_1\(\frac{w_n}{z_m} z'\) \frac{\exp \(- \frac{w_n}{z_m} |\vec x - \vec x'|\)}{ |\vec x - \vec x'|}~~,
\end{eqnarray}
where the normalization constants $N_n^v$ and $N_n^A$ are given by
\begin{eqnarray}
N^v_n \!\!\!&=&\!\!\! \frac{z_m^2}{2}\left\{ J_1'(x_n)^2 +\left( 1- \frac{1}{x_n^2}\right) J_1(x_n)^2  \right\} , \\  N^A_n \!\!\!&=&\!\!\! \frac{z_m^2}{2}   J_2(w_n)^2  ~.
\end{eqnarray}
Putting these Green's functions into Eq.~(\ref{correction}), we can compute perturbative corrections to the leading order solutions obtained in Sec. \ref{SBOHW}.

\end{document}